\documentclass[12pt]{article}
\usepackage{a4wide}

\newcommand{\beq}{\begin{equation}}
\newcommand{\eeq}{\end{equation}}
\newcommand{\bea}{\begin{eqnarray}}
\newcommand{\eea}{\end{eqnarray}}

\newcommand{\pa}{\partial}
\newcommand{\bma}[1]{\mbox{\boldmath${#1}\/$}}
\newcommand{\w}{\omega}
\newcommand{\dir}{\bma{\Omega}}

\begin{document}

{\Large
\centerline{{\huge {\bf Stochastic backgrounds of gravitational }}}
\centerline{{\huge {\bf waves and spherical detectors          }}}}
\vskip 1.1cm

\centerline{{\Large J.\ Alberto Lobo$^{(1)}$ and \'Alvaro Montero$^{(2)}$}}

\vskip  0.4cm
\centerline{{\Large Department de F\'{\i}sica Fonamental}}
\vskip  0.05cm
\centerline{{\Large Universitat de Barcelona}}
\vskip  0.05cm
\centerline{{\Large Diagonal 647, Barcelona 08028}}
\vskip  0.05cm
\centerline{{\Large Spain}}
\vskip 0.8cm
\centerline{{\Large (1) e-mail: lobo@ffn.ub.es}}
\vskip  0.05cm
\centerline{{\Large (2) e-mail: montero@ffn.ub.es}}
\vskip  0.15cm

\begin{center}
{\bf ABSTRACT}
\end{center}

\noindent
The analysis of how a stochastic background of gravitational radiation
interacts with a spherical detector is given in detail, which leads to
explicit expressions for the system response functions, as well as for
the cross-correlation matrix of different readout channels. It is shown
that distinctive features of GW induced random detector excitations,
relative to locally generated noise, are in practice insufficient to
separate the signal from the noise by means of a \emph{single} sphere,
if prior knowledge on the GW spectral density is nil. The situation
significantly improves when such previous knowledge is available, due
to the omnidirectionality and multimode capacities of a spherical GW
antenna.


\newpage

\section{Introduction \label{sec.1}}

Stochastic backgrounds of Gravitational Waves (GWs) are amongst the
most interesting sources to be detected by the upcoming new generation
of antennas: such kind of gravitational radiation will convey to us
e.g.\ information on the structure of the Universe at its earliest
evolutionary stages. This is quite unique to gravitational waves, 
due to their extremely weak coupling to other forms of matter and 
radiation, and will therefore greatly enhance our understanding of 
Cosmology as well as Fundamental Physics. Literature on the possible
forms, origins and spectral shapes of various gravitational wave
backgrounds is rich ---see for example~\cite{Maggiore} for a detailed
review and further references on the subject.

The stochastic nature of gravitational background signals, together
with the same stochastic origin of the local noise generated in the
detector itself, makes very difficult to tell local noise from actual
signal. Therefore, detection strategies to extract that kind of
gravitational signals from detector data are based on finding
\emph{differential properties} capable to distinguish between
both processes.

The standard procedure to identify a stochastic signal is to use \emph{two}
(or more) antennas, and cross-correlate their outputs. The concept is that,
while local noises in separate detectors are uncorrelated, the stochastic
signal is common to both. Signal to noise ratio therefore builds up on
the basis of long-term average compensation of local noise fluctuations,
and can be seen to develop as $T^{1/4}$, where $T\/$ is the integration
time~\cite{bfs91}. Specific strategies for GW background detection have
been considered for two bar detectors~\cite{pia}, two
interferometers~\cite{allen}, a bar and an interferometer~\cite{als}
and two spherical detectors~\cite{vitale}.

A spherical GW detector is a \emph{multimode} device which, in a number
of senses, behaves like an array of bars~\cite{massimo,jm93,clo}. This
suggests that one might try and take advantage of such multimode capacity
to perform a search for a GW background by suitably cross-correlating
the different detector readouts. A procedure like this would therefore
enable the determination of the GW background spectral density with a
\emph{single} detector ---a significant advantage to rid the usual
method (sketched above) of inherent uncertainties bound to the fact
that no two real detectors are exactly identical.

One might \emph{a priori} expect in this direction that specific
\emph{signal correlation patterns} happen between pairs of sensor
readouts which be \emph{not} shared by the local sources of noise,
thereby making possible to filter the latter out. This paper is
concerned with the analysis of the possibilities offered by the
sphere, as a multimode device, for the detection of a stochastic
background of GWs, defined by a \emph{spectral density} function.

As we shall see, though, local noise and GW background random amplitudes
actually show no distinctive correlation pattern in the output channels,
except for the obvious fact that GWs are only seen in quadrupole (perhaps
also monopole) antenna modes~\cite{lobo}. So, in the absence of some kind
of previous information on the functional form of the GW spectrum, the
above expectations fail. If, on the other hand, spectral information
is available ahead of time (\emph{templates}) then the sphere has a
very good performance in its frequency band of sensitivity. This can
improve efficiency up to an order of magnitude in energy compared with
single readout GW detectors (bars and interferometers), due to the
sphere being both \emph{omnidirectional} and \emph{multimode}, as
we shall see.

The outline of the article will be the following. In section~\ref{sec.2}
we set up definitions and notation for the stochastic background of GWs,
and in section~\ref{sec.3} we derive the sphere's response to this
signal. In section~\ref{sec.corr} we present the correlation functions
between the sphere's responses induced by the gravitational signal at 
two points on its surface, where motion sensors will be attached. 
Section~\ref{sec.6} is dedicated to calculate the same correlation function presented
in section~\ref{sec.corr} but in this case induced by local noise, and discusses, in 
the light of these results, various possible strategies to retrieve information
about the incoming gravitational signal spectrum with a \emph{single}
spherical detector. We close with a summary of conclusions in
section~\ref{sec.8}, and also include a brief mathematical appendix.

\section{The stochastic signal \label{sec.2}}

We give in this section a few definitions and notation for a stochastic
background of gravitational radiation, and its characterization through
the spectral density $S_h(\w)$.

A background of gravitational waves can be expressed as a linear superposition 
of plane waves coming from all directions and with all possible polarizations,

\beq
 h_{ij} (t,{\bf x}) = \int_{-\infty}^{\infty}\frac{d\w}{2\pi}
 \int d^2\Omega\hspace{0.1 cm} e^{-i\w(t-\dir\cdot{\bf x})}
 \int_0^{2 \pi} d\psi \hspace{0.1 cm} \sum_{A}
 \tilde{h}_A(\w;\dir,\psi) \hspace{0.1 cm}
 G^A_{ij}(\dir,\psi)
 \label{eq:wavegen} 
\eeq
where $\dir$ is a unit vector pointing to a generic location, $\psi\/$
is a rotation angle around this vector, $\tilde{h}_A(\w;\dir,\psi)$ are
the wave amplitudes and $G^A_{ij}(\dir,\psi)$ the polarization matrices.
General Relativity predicts only \emph{two transverse} polarization modes.
Nevertheless, another four degrees of freedom must generally be allowed
for. We thus have altogether \emph{six} independent $3\!\times\!3$
symmetric polarization matrices. If we choose a right-handed triad of
orthonormal vectors \bma{m}, \bma{n} and $\dir$, where $\dir$ is aligned 
with the propagation direction, a suitable parametrization for the
$G\/$-matrices is the following:

\bea
& &  G^{M}_{ij}(\dir,\psi) = \sqrt{\frac{2}{3}} \hspace{0.2 cm} \delta_{ij}  
     \hspace{1.9 cm} , \hspace{0.5 cm} 
     G^{L_0}_{ij}(\dir,\psi) = \sqrt{\frac{1}{3}}\,\left( 3 \Omega_i \Omega_j  -\delta_{ij}\right) 
     \nonumber \\
& &  G^{L_{m}}_{ij}(\dir,\psi) = m_i \Omega_j + m_j \Omega_i  
     \hspace{0.7 cm} , \hspace{0.5cm}
     G^{L_{n}}_{ij}(\dir,\psi) = n_i \Omega_j + n_j \Omega_i \nonumber \\
& &  G_{ij}^{T_{\times}}(\dir,\psi) = m_i n_j + n_i m_j  
     \hspace{0.85 cm} , \hspace{0.5 cm}
     G^{T_+}_{ij}(\dir,\psi) = m_i m_j - n_i n_j 
\eea
where we give to the index A in $G^A_{ij}(\dir,\psi)$ the following
meanings: $\!A=\!M$ is the \emph{monopole} component, $A\!=\!L_0,L_m,L_n$
are \emph{longitudinal quadrupole} components, and $A\!=\!T_+,T_{\times}$
are the two \emph{transverse quadrupole} components. These matrices
satisfy normalization conditions

\beq
 \sum_{ij}G^A_{ij}G^B_{ji}=2\delta^{AB}
 \label{eq:2}
\eeq

If $(\theta,\phi,\psi)$ are Euler angles relating the laboratory frame to
the just described wave frame then

\bea
& & \bma{m} = (\cos\psi\cos\theta\cos\phi-\sin\psi\sin\phi \hspace{0.1 cm},
               \cos\psi\cos\theta\sin\phi+\sin\psi\cos\phi \hspace{0.1 cm},
	      -\cos \psi \sin \theta) \nonumber \\
& & \bma{n} = (-\sin\psi\cos\theta\cos\phi-\cos\psi\sin\phi \hspace{0.1 cm},
               -\sin\psi\cos\theta\sin\phi+\cos\psi\cos\phi \hspace{0.1 cm},
                \sin\psi\sin\theta) \nonumber \\   
& & \bma{\Omega} = (\sin\theta\cos\phi,\sin\theta\sin\phi, \cos\theta)
\eea
where we can see that a choice for the angles $\theta,\phi$ gives a
direction in space and a choice for $\psi$ gives the definition of the
two transverse polarization states for a particular direction.

The ``\emph{electric}'' components $R_{0i0j}(t)$ of the Riemann tensor
at the center of mass of the sphere are

\beq
 R_{0i0j}(t) = \frac{1}{2} h_{ij,00}({\bf 0},t) = -
 \int_{-\infty}^{\infty} d\w \hspace{0.1 cm} \w^2e^{-i\w t}
 \int d^2\Omega\int d\psi\;\sum_{A=0}^5
 \tilde{h}_A(\w;\dir,\psi)\;G^A_{ij}(\dir,\psi) ,
 \label{eq.riem}
\eeq
and these are the only components carrying information on an incoming 
GW~\cite{el73}.

In this paper we shall be considering an unpolarized stochastic bath of
GWs which is also isotropic and stationary. Its statistical properties
are therefore encoded in its \emph{power spectrum} function, defined by
the following equation:

\beq
 \langle\tilde{h}_A^*(\omega,\dir,\psi)\,
 \tilde{h}_{A'}(\omega',\dir',\psi')\rangle = 
 \delta(\omega-\omega')\,\left\{\frac{1}{4\pi}\delta^2(\dir,\dir')\,
 \frac{1}{2 \pi}\delta (\psi-\psi')\right \}\,
 \left\{\delta_{AA'}\,\frac{1}{2}S_h(\omega)\right\} 
\label{eq:ensaver}
\eeq
where $\langle -\rangle$ stands for \emph{ensemble average}.
There are alternative ways to characterize the frequency spectrum, for
example, through an energy density per unit logarithmic interval of
frequency, or in terms of a characteristic amplitude of the stochastic
background. The definitions for such quantities, as well as their
mutual relationships can be found in reference~\cite{Maggiore}.

\section{The sphere's response \label{sec.3}}

In this section we calculate the sphere's response to the background of
gravitational waves given by equation~(\ref{eq:wavegen}). We shall use
henceforth the general formalism and notation of reference~\cite{lobo}.

We assume that the sphere's response ${\bf u}({\bf x},t)$ to an incoming
signal is the solution to the partial differential equation

\beq
 \varrho\frac{\pa^2{\bf u}}{\pa t^2}-\underbrace{\left\{ \mu\,\nabla^2{\bf u}
 + (\lambda + \mu)\,\nabla(\nabla\cdot{\bf u})\right\}}_{{\cal L}{\bf u}} =
 {\bf f}({\bf x},t)
 \label{eq:diff}
\eeq
with suitable initial and boundary conditions. Here, $\lambda$ and $\mu$
are the material's elastic Lam\'e coefficients~\cite{ll70}, and $\varrho$
is the sphere's density. In the rhs of equation~(\ref{eq:diff}),
f$_i({\bf x},t)=\varrho c^2 R_{0i0j}(t)x_j$ is the \emph{density} of
gravitational wave \emph{tidal} forces, ensuing from the \emph{geodesic
deviation} equation ---see e.g.~\cite{wald}. The latter term can be
split up into its monopole and quadrupole components according
to~\cite{lobo}

\begin{equation}
 {\bf f}({\bf x},t) = {\bf f}^{(00)}({\bf x})\,g^{(00)}(t)\ +\ 
 \sum_{m=-2}^2\,{\bf f}^{(2m)}({\bf x})\,g^{(2m)}(t)
 \label{eq.ddf}
\end{equation}
with

\bea
& & {\rm f}_i^{(00)}({\bf x}) = \varrho\,E_{ij}^{(00)}\,x_j \hspace{0.7cm},
    \hspace{0.5cm} g^{(00)}(t) =
    \frac{4\pi}{3}\,E_{ij}^{*(00)}\,R_{0i0j}(t)\,c^2
 \nonumber \\
& & {\rm f}_i^{(2m)}({\bf x}) = \varrho\,E_{ij}^{(2m)}\,x_j \hspace{0.5 cm},
    \hspace{0.5 cm} g^{(2m)}(t) =
    \frac{8\pi}{15}\,E_{ij}^{*(2m)}\,R_{0i0j}(t)\,c^2
    \qquad (m = -2,\ldots,2)
\label{eq.monqua}
\eea

The definition for the matrices $E_{ij}^{(00)}$ and $E_{ij}^{(2m)}$ can be
found in appendix A. The generic response function ${\bf u}({\bf x},t)$ can
be expressed as an orthogonal series expansion, which only involves
quadrupole and monopole terms:

\begin{equation}
   {\bf u}({\bf x},t) = \sum_{n=1}^\infty\,\frac{a_{n0}}{\omega_{n0}}
   \,{\bf u}_{n00}({\bf x})\,g_{n0}^{(00)}(t)\ +\ 
   \sum_{n=1}^\infty\,\frac{a_{n2}}{\omega_{n2}}\,\left[\sum_{m=-2}^2
   \,{\bf u}_{n2m}({\bf x})\,g_{n2}^{(2m)}(t)\right]
   \label{eq.eigexp}
\end{equation}
where ${\bf u}_{nlm}$ are \emph{spheroidal} wavefunctions corresponding
to the ($2l+1$)-degenerate eigenfrequency $\w_{nl}$, i.e.,

\beq
 {\cal L}{\bf u}_{nlm}({\bf x}) = -\omega^2_{nl}\varrho{\bf u}_{nlm}({\bf x})
\eeq
with ${\cal L}$ the differential operator defined in equation~(\ref{eq:diff});
$a_{n0}$ and $a_{n2}$ are projection coefficients ---see~\cite{mnras} for
details on definitions and values of these quantities.

Finally, $g_{nl}^{(lm)}(t)$ are \emph{convolution products} between the
driving terms $g^{(lm)}(t)$ and the corresponding eigenmode oscillation
function ---in this case represented by an ideally non-dissipating, purely
sinusoidal vibration:

\begin{equation}
 g_{nl}^{(lm)}(t)\equiv\int_0^t dt'\hspace{0.1 cm}  g^{(lm)}(t')\,
                 \sin\left\{\omega_{nl}(t-t')\right\} \hspace{0.5 cm}
 (l=0,2 \hspace{0.2 cm} ; \hspace{0.1 cm} m=-l,...,l)
 \hspace{0.5 cm}
 \label{eq.gnl}
\end{equation}

We shall later relax this ideal behavior hypothesis in order to cope with
the stationary state the system will eventually reach under the action of
long term random excitations.

We next calculate the sphere's response to the background of gravitational
waves given by equation (\ref{eq:wavegen}). The only quantities we have
to evaluate are the functions $g^{(lm)}_{nl}(t)$ using
equations~(\ref{eq.gnl}) and~(\ref{eq.monqua}), with the components of
the Riemann tensor given by equation~(\ref{eq.riem}). The following is
readily found:

\beq
 g^{(lm)}_{nl}(t)=-\int_{-\infty}^{\infty} d\w
 \hspace{0.1 cm} \w^2\,T(t;\w,\w_{nl})
 \,\int d^2\Omega\,d\psi\;\sum_A\tilde{h}_A(\w;\dir,\psi)
 \;G_{ij}^A(\dir,\psi)\,E_{ij}^{*(lm)}
 \label{eq.glmsb}
\eeq
where

\beq
 T(t;\w,\w_{nl})\equiv\int_0^t dt' \hspace{0.1 cm}
 e^{-i\w t'}\,\sin\left\{\omega_{nl}(t-t')\right\} 
 \label{eq.T}
\eeq
is a function strongly peaked at $\w=\pm\w_{nl}\/$ for large values
of $t$. The sphere's response ${\bf u}({\bf x},t)$ is then given by
equation~(\ref{eq.eigexp}) with $g^{(lm)}_{nl}(t)$ given by
equation~(\ref{eq.glmsb}).

In actual practice, the sphere's motions are sensed at a number of
different positions on its surface, where \emph{motion sensors} are
attached. We consider devices which are only sensitive to radial
displacements, and attached to the sphere's surface at positions

\beq
 {\bf x}_a = R\,{\bf n}_a\ ,\qquad a=1,\ldots,J
 \label{eq.sensors}
\eeq
where $R\/$ is the radius of the sphere, and ${\bf n}_a$ a unit vector
pointing outward at the $a\/$-th position. There will consequently be
$J\/$ readout channels with output displacements

\beq
 u_a(t)\equiv {\bf n}_a\cdot {\bf u}({\bf x}_a,t)\ ,\qquad a=1,\ldots,J
 \label{eq.channel}
\eeq

If equation~(\ref{eq.eigexp}) is now used, these are seen to be

\bea
 u_a^{GW}(t) & = & \sum_{n=1}^{\infty} \frac{a_{n0}}{\w_{n0}}
 \hspace{0.1 cm} A_{n0}(R) \hspace{0.1 cm} Y_{00}({\bf n}_a)
 \hspace{0.1 cm} g_{n0}^{(00)}(t)  
 \hspace{0.3 cm} + \nonumber \\
 & + & \sum_{n=1}^{\infty} \frac{a_{n2}}{\w_{n2}}\;A_{n2}(R)
 \left[\,\sum_{m=-2}^{2}
 Y_{2m}({\bf n}_a) \hspace{0.1 cm} g_{n2}^{(2m)}(t)\right]\ ,
 \ \ \ a=1,\ldots,J
\label{eq.output}
\eea
where $Y_{lm}$ are spherical harmonics, and $A_{nl}(R)$ is a radial function
coefficient in the spheroidal wavefunction ${\bf u}_{nlm}({\bf x})$; this
one depends on whether we are dealing with a solid or a hollow sphere
---see~\cite{lobo} or~\cite{vega} for each case, respectively.

Because the time dependent terms $g_{n2}^{(2m)}(t)$ in~(\ref{eq.output})
are \emph{random} in nature, the readouts $u_a(t)$ are consequently random,
too. In addition, the linear character of the relationship between both
ensures that certain properties of the driving terms, such as e.g.\ 
gaussianity, directly carry over to the outputs. We now propose to
investigate the relevant statistical properties of the latter.

\section{Cross-correlation functions \label{sec.corr}}

As already stated in the Introduction, an essential ingredient in our
analysis are the \emph{cross-correlations} between different output
channels. In order to make them useful, however, we first need to
establish their \emph{theoretical} structure, i.e., to assess their
behavior under ideal conditions of infinite integration times, and
relate them to the unique characteristic function $S_h(\omega)$ ---the
signal spectral density.

Correlations between the outputs at the readout channels are naturally
defined by

\beq
 R_{ab}(t,\tau)\equiv\langle u_a^*(t)\,u_b(t+\tau)\rangle  
\label{eq.corfun}
\eeq
where $\langle-\rangle$ stands for ensemble average again. We must now
introduce the function $u_a^{GW}(t)$, given by equation~(\ref{eq.output}),
into equation~(\ref{eq.corfun}) to calculate the correlation functions we
are looking for. When averages are taken, equation~(\ref{eq:ensaver}) for
the ensemble average of the wave amplitudes comes into play. Dirac
$\delta$-functions appearing in that equation allow us to easily perform
the integrals in $\dir'$, $\psi'$ and $\omega'$. The remaining integrals
in $\dir$, $\psi$ can then be done after evaluating

\beq
 G_{ijkl}\equiv\int\,\frac{d\psi}{2\pi}\;\int\,\frac{d^2\Omega}{4\pi}
 \;\sum_A\,e_{ij}^A(\dir,\psi)\,e_{kl}^A(\dir,\psi)
 \label{eq.gijkl}
\eeq
which yields the following result:

\beq
 G_{ijkl} = \underbrace{\frac{2}{3}\,\delta_{ij}\delta_{kl}}_{monopole} +
            \underbrace{2N\,\left[-\frac{1}{15}\,\delta_{ij}\delta_{kl} +
            \frac{1}{10}\,(\delta_{ik}\delta_{jl}+
            \delta_{il}\delta_{jk})\right]}_{quadrupole}
 \label{eq.gijklsol}
\eeq

The index $A$ in equation (\ref{eq.gijkl}) runs over the polarization
modes of the incoming gravitational signal. And the result given in
equation~(\ref{eq.gijklsol}) takes into account the monopole mode and
$N\/$ quadrupole modes (note that $N\leq 5$, for example $N=2$ in
General Relativity, and that all of them have the same contribution
to $G_{ijkl}$).

It is now readily seen that there is no cross-correlation between
monopole and quadru\-pole modes, and that the correlation functions
between the outputs induced by the gravitational signal split up as

\beq
 R_{ab}^{GW}(t,\tau) = R^{(0)GW}_{ab}(t,\tau) + R^{(2)GW}_{ab}(t,\tau)
 \label{eq.corrgw}
\eeq
where $R^{(0)GW}_{ab}$ and $R^{(2)GW}_{ab}$ are monopole and quadrupole
terms, respectively, and have the form

\bea
 R^{(0)GW}_{ab}(t,\tau) & = & \hspace{0.28 cm} 2
 \hspace{0.28 cm} P_0(\dir_a\!\cdot\dir_b)\, 
 \sum_{n,n'=1}^{\infty}\,
 \frac{a_{n0}a_{n'0}\,A_{n0}(R)A_{n'0}(R)}{\omega_{n0}\omega_{n'0}}\,
 f_{nn'}^{(0)}(t,\tau) \nonumber \\
 R^{(2)GW}_{ab}(t,\tau) & = & \frac{4N}{15}\,P_2(\dir_a\!\cdot\dir_b)\,
 \sum_{n,n'=1}^{\infty}\,
 \frac{a_{n2}a_{n'2}\,A_{n2}(R)A_{n'2}(R)}{\omega_{n2}\omega_{n'2}}\,
 f_{nn'}^{(2)}(t,\tau) \label{eq.corrgwl}
\eea
with $P_0$ the zero-th Legendre polynomial ---which is identically equal
to~1 for any value of its argument, but we keep it explicit for the sake
of structure clarity---, $P_2$ is the second Legendre polynomial, and

\beq
 f_{nn'}^{(l)}(t,\tau)\equiv\int_0^{\infty}\,\frac{d\omega}{2 \pi}
 \;\omega^4\,{R}{\rm e}\left[
 T^*(t,\omega,\omega_{nl})\,T(t+\tau,\omega,\omega_{n'l})\right]
 \,S_h(\omega)\ ,\qquad l=0,2
 \label{eq.4-7}
\eeq

Here we see how the spectral density $S_h(\omega)$ characterizing the
incoming signal appears, and its relation to the correlation functions.

In a stationary random process, the second order correlation functions
are time-shift invariant (by definition), i.e., if we use the
definition~(\ref{eq.corfun}) then the function $R\/$ should \emph{only}
depend on its second argument, $\tau$. Time-shift invariance has been
assumed for the exciting GW bath, as we have characterized it by the spectral
density $S_h(\omega)$. Nevertheless, the sphere's response is calculated
with the series expansion~(\ref{eq.eigexp}) in which a choice for the
state of the sphere at $t=0$ (initial conditions) is assumed which,
therefore, breaks that invariance.

The correlation functions we are looking for can actually be obtained
from~(\ref{eq.corrgw}) and~(\ref{eq.corrgwl}) by taking their limit when
$t\/$ approaches infinity, because in this limit the property of time
shift invariance is recovered, as the system looses memory of any
particular initial conditions in the remote past.

Memory is however not lost in the kind of ideal, non-dissipative elastic
body we have described in the preceding sections. Meaningful results
can only be obtained if some kind of dissipation, no matter how large
or small, is present in the system. This is therefore the appropriate
place to introduce dissipation, and we shall do it in the standard way
of assuming an exponentially decaying amplitude for the oscillation
eigenmodes, i.e., we make the replacement

\beq
 \sin(\omega_{nl}\tau)\ \ \longrightarrow\ \ 
 e^{-\gamma_{nl}\tau}\,\sin(\omega_{nl}\tau)
\eeq
where $\gamma_{nl}$ is proportional to the inverse of the decay time of
the associated mode, which is in all cases of interest much larger than
the period of oscillation of that mode, or

\beq
 \gamma_{nl}\ll\w_{nl}
\eeq

If we express the limits of $R^{(0)GW}_{ab}(t,\tau)$ and
$R^{(2)GW}_{ab}(t,\tau)$ by dropping the first argument
then the following results are readily obtained:

\bea
 & & R^{(0)GW}_{ab}(\tau) = \hspace{0.28 cm} 2 \hspace{0.28 cm}
 P_0(\dir_a\!\cdot\dir_b)\,
 \int_{-\infty}^{\infty} \frac{d\omega}{2\pi} \hspace{0.2 cm}
 \tilde{R}^{(0)}(\omega)\,\frac{1}{2} S_h(\omega)\,e^{i\omega\tau}
 \label{eq.corrfnm} \\[1 em]
& &  R^{(2)GW}_{ab}(\tau) =  \frac{4N}{15}\,P_2(\dir_a\!\cdot\dir_b) 
 \int_{-\infty}^{\infty} \frac{d\omega}{2\pi} \hspace{0.2 cm} 
 \tilde{R}^{(2)}(\omega)\,\frac{1}{2} S_h(\omega)\,e^{i\omega\tau}
 \label{eq.corrfnq}
\eea
where

\beq
 \tilde{R}^{(l)}(\omega) = \sum_n\;[a_{nl} A_{nl}(R)]^2\;
 \w^4\left|L_{nl}(\w)\right|^2
 \label{eq.transfn}
\eeq
and $L_{nl}(\w)$ is a Lorentzian curve:

\beq
 L_{nl}(\w)\equiv\frac{1}{\w^2-\w_{nl}^2 + 2i\gamma_{nl}\,\w}
 \label{loren}
\eeq

Equations~(\ref{eq.corrfnm}) and~(\ref{eq.corrfnq}) have a foreseeable
\emph{functional} structure: the correlation functions are Fourier
transforms of the spectral density of the driving stochastic signal
times the sphere's transfer function $\tilde{R}^{(l)}(\w)$, and
this is a characteristic sum of Lorentzian curves centered on the
resonant frequencies, and with appropriate weights. Note also that
different harmonics ($n,n'$) appear to be uncorrelated.

\subsection{Mode channels \label{sec.5}}

The \emph{algebraic} structure of the correlation function \emph{matrix}
requires some further consideration. For $P_2(\dir_a\!\cdot\dir_b)$ is
a symmetric matrix whose rank\footnote{
Actually, {\sf rank}$\{P_l(\dir_a\!\cdot\dir_b)\}=2l+1$, provided no
two of the $J\/$ vectors $\dir_a$ are parallel, and provided of course
that $J\geq 2l+1$ ---see \protect\cite{mnras} for full details.} is at
most~5, which means one can form~5 linear combinations

\beq
 y^{(m)}(t) = \sum_{a=1}^J\,v_a^{(m)}\,u_a(t)\ ,\qquad
 m=-2,\ldots,2
 \label{modech}
\eeq
where the coefficient $v_a^{(m)}$ is the $a\/$-th component of the
(normalized) eigenvector associated to the $m\/$-th non-null eigenvalue
of $P_2(\dir_a\!\cdot\dir_b)$, $\zeta_m^2$, say, which is always positive.

If we now define a new correlation function matrix

\beq
 {\cal R}_{mm'}^{GW}(\tau)\equiv\langle y^{(m)*}(t)\,y^{(m')}(t+\tau)\rangle\ ,
 \qquad m,m'=-2,\ldots,2
 \label{modecorr}
\eeq
we readily see that

\beq
 {\cal R}_{mm'}^{GW}(\tau) = \frac{4N}{15}\,\zeta_m^2\,\delta_{mm'}\;
 \int_{-\infty}^{\infty}\;\frac{d\omega}{2\pi}\hspace{0.2 cm} 
 \tilde{R}^{(2)}(\omega)\,\frac{1}{2}S_h(\omega)\,e^{i\omega\tau}
 \label{diacorr}
\eeq

Thus the 5 quadrupole channels $y^{(m)}(t)$ are \emph{uncorrelated} with
one another. They are therefore particularly well suited for improved
estimation of the \emph{spectral density} $S_h(\omega)$, since the latter
is common to all of them. One may recall here that $y^{(m)}(t)$ are
actually \emph{mode channels} for the well known {\sl TIGA\/}~\cite{jm95}
and {\sl PHC\/}~\cite{ls} sensor distributions, in which cases

\beq
 v_a^{(m)}\propto Y_{2m}(\dir_a)\ \ \ \ \mbox{for {\sl TIGA\/} and
 {\sl PHC}}
\eeq

Note that the fact that the $y^{(m)}(t)$ are uncorrelated is independent
of whether the motion sensors are resonant or not, so it applies to a
system like the recently proposed \emph{dual sphere}~\cite{dual}, too.
Note also that sensor geometries such as {\sl TIGA\/} and {\sl PHC},
for which the $y^{(m)}(t)$ are \emph{mode channels} ---i.e., quantitites
whose frequency spectrum $\tilde y^{(m)}(\w)$ is directly proportional to
that of the GW quadrupole amplitudes $\tilde g^{(2m)}(\w)$---, allow for
a very clear and direct physical interpretation of the correlation
coefficients~(\ref{diacorr}). They are therefore also preferred from
the specific point of view of the signals considered in this paper.

Summing up, the main result so far is that cross-correlations between
different output channels of a spherical GW detector, whether direct
sensor readouts or quadrupole (mode) channels, possess a specific
\emph{pattern}, which is given by equations~(\ref{eq.corrfnq})
or~(\ref{diacorr}), respectively. The problem is of course that
a random signal must be told from also random \emph{local noise};
we thus review next the relevant properties of the latter.

\section{Local noise and detection strategies \label{sec.6}}

We study in this section the \emph{multipole} characteristics of the
correlation functions between the outputs at two points of the sphere
induced by the local noise. 

The actual readouts of the spherical antenna are in fact superpositions
of those induced by the GW and by the local noise, i.e.,

\beq
 u_a(t) = u_a^{GW}(t) + u_a^{LN}(t)
 \label{eq:output}
\eeq

We shall generically consider that

\beq
 u_a^{LN}(t) = \sum_{n,l,m}\,q_{nlm}(t)\,A_{nl}(R)\,Y_{lm}({\bf n}_a)
 \label{eq.outputLN}
\eeq
where $q_{nlm}(t)$ are \emph{narrow band} stochastic time series, with
correlation times of order $1/\gamma_{nl}$. Because these $q_{nlm}(t)$
are the amplitudes of different \emph{normal modes} of the solid, they
are statistically \emph{uncorrelated}, or

\beq
 \langle q_{nlm}^*(t)\,q_{n'l'm'}(t+\tau)\rangle =
 \delta_{nn'}\,\delta_{ll'}\,\delta_{mm'}\;{\cal F}_{nl}(\tau)
 \label{eq.ensavef}
\eeq
with,

\beq
 {\cal F}_{nl}(\tau) = \int_{-\infty}^{\infty}\;\frac{d\w}{2\pi}\,\;
 |L_{nl}(\w)|^2\;\;\frac{1}{2}S_{LN}(\w)\,e^{i\w\tau}
 \label{eq.oscill}
\eeq
where $L_{nl}(\w)$ is given by~(\ref{loren}), and $S_{LN}(\w)$ is the
input spectral density of local detector noise. We need not go into
the details of the specific form of this function here, which is a
superposition of thermal and transducer back-action noise ---see~\cite{dual}.

The system readout correlation functions of local noise, defined by

\beq
 R_{ab}^{LN}(\tau)\equiv\langle u_a^{LN*}(t)\,u_b^{LN}(t+\tau)\rangle
\eeq
are therefore given by

\beq
 R^{LN}_{ab}(\tau) = \sum_{l}\;R^{(l)LN}_{ab}(\tau)
 \label{eq.corrfLNs}
\eeq
where

\beq
 R^{(l)LN}_{ab}(\tau) = \frac{2l+1}{4 \pi}\;P_l(\dir_a\!\cdot\dir_b)
 \;\int_{-\infty}^{\infty}\;\frac{d\omega}{2\pi}
 \;\;\sum_n |A_{nl}(R)|^2\,|L_{nl}(\w)|^2
 \ \frac{1}{2}\,S_{LN}(\omega)\;e^{i\omega\tau}
 \label{eq.corrfLN}
\eeq

These expressions show that local noise correlations split up in a
way similar to~(\ref{eq.corrgw}), except of course that we now have
contributions from \emph{all} multipoles rather than just monopole
and quadrupole modes, characteristic of GW signals. In particular,
we can easily assess how do quadrupole channels ---as defined
in~(\ref{modech})--- correlate with one another in a local noise
dominated detector:

\bea
 \langle y_{LN}^{(m)*}(t)\,y_{LN}^{(m)}(t+\tau)\rangle & = & \nonumber \\[1 em]
 & & \hspace*{-4 em}  \frac{5}{4 \pi}\,\zeta_m^2\,\delta_{mm'}
 \;\int_{-\infty}^{\infty}\;\frac{d\w}{2\pi}
 \;\;\sum_n |A_{n2}(R)|^2\,|L_{n2}(\w)|^2
 \ \frac{1}{2}\,S_{LN}(\w)\;e^{i\w\tau} \nonumber \\[1 em]
 & & \hspace*{-7.2 em} +\ \sum_{l\neq 2}\;\frac{2l+1}{4 \pi}
 \;{\cal P}_{mm'}^{(l)}\;\int_{-\infty}^{\infty}\;\frac{d\w}{2\pi}
 \;\;\sum_n |A_{nl}(R)|^2\,|L_{nl}(\w)|^2
 \ \frac{1}{2}\,S_{LN}(\w)\;e^{i\w\tau} \label{eq.41}
\eea
where

\beq
 {\cal P}_{mm'}^{(l)} = \sum_{a,b}\,v_a^{(m)*}v_b^{(m')}\,
 P_l(\dir_a\!\cdot\dir_b)\ ,\qquad m=-2,\ldots,2
\eeq
is generally a non-diagonal matrix\footnote{
Exceptionally, it is diagonal for certain values of $l\/$ in certain sensor
configurations ---see a discussion in~\cite{sl}.}.

The important conclusion to be drawn from equations~(\ref{eq.corrfLN})
and~(\ref{eq.41}) is that monopole and quadrupole correlations follow
precisely the \emph{same algebraic pattern}, whether they are dominated
by local noise or by a bath of random GWs. Extra terms, such as those
in the last line of~(\ref{eq.41}), do not really provide a distinctive
feature of local noise, for they are only significant at frequencies
different from those of the quadrupole modes, as determined by the
presence of the frequency dependent coefficients $L_{nl}(\w)$ in
each case.

This is a very unfortunate circumstance, indeed, for it in principle
renders useless the cross-correlation algorithm between different
detector channels as an efficient means of filtering out local noise.
Are there expedient alternatives?

\subsection{Filtering strategies \label{sec.7}}

There are two interesting possibilities one can try to filter out local
noise effects with a \emph{single} multimode detector:

\begin{itemize}
\item[\bf i)] Since different normal modes are statistically uncorrelated,
one is tempted to consider e.g.\ the first and second quadrupole modes
of a sphere~\cite{clo,vega,dual} as a pair of independent antennas, then
use cross-correlation between them to filter local noise~out.

This however cannot possibly work because, even if local noise induced
fluctuations are independent in either mode, so are also GW induced
fluctuations. And, as we have just seen, there is no distinctive
feature as to how noise gets added to the signal at different quadrupole
modes. Actually, as seen in equations~(\ref{modecorr}) and~(\ref{eq.41}),
there is no way to tell which part in $y^{(m)}(t)$ is due to GWs and
which is due to local noise.

\item[\bf ii)] The fundamental difference between a bath of random GWs and
local noise is that, while the former can only affect \emph{quadrupole}
modes (or, at most, also monopole modes), local noise excites \emph{all}
modes, instead ---not just quadrupole/monopole. Given that, in a
realistic GW antenna, mode linewidths are extremely narrow, i.e.,
energy transfer between different modes is extremely slow, an attractive
possibility to use a \emph{single} sphere as a detector of a
\emph{continuous} flow of stochastic GWs would be to identify the
effect of the latter as a \emph{temperature excess} in the quadrupole
modes, relative to other modes ---dipole, octupole, etc.

While this may look like a reasonable approach, it is in fact an
impractical one, as we can appreciate by the following argument.
The temperature of a mode is of course a measure of the variance
of its random oscillations; since these are a \emph{narrow band}
stochastic process, their correlation time, $\tau_{\rm corr}$,
is quite long, hence a reliable estimate of the variance requires
taking data during a suitable number of correlation times for
averaging. If $t_{\rm obs}$ is such time then the \emph{relative}
error in the temperature estimation is of the order of

\beq
 \frac{\delta T}{T}\simeq\sqrt{\frac{\tau_{\rm corr}}{t_{\rm obs}}}
 \label{vari}
\eeq

Let us consider a few likely numbers: an SQL detector ---such as described
e.g.\ in~\cite{dual}--- should be capable of sensing energy innovations
of magnitude $kT/Q$, where $k\/$ is Boltzmann's constant, $T\/$ is the
temperature of the mode (e.g.\ the first quadrupole), and $Q\/$ its
mechanical quality factor. An optimized device should be able to reach
$T/Q\sim 10^{-8}$\,K$^{-1}$, for example with $Q=10^7$ and
$T=100$\,milli-Kelvin. A spherical antenna whose considered resonance
happens at 1\,kHz, say, thus has a correlation time of $2\pi Q/\w_{12}=10^4$
seconds. On the other hand, the precision required in the measurement of
the mode temperature to match the SQL is $\delta T/T=10^{-7}$ for the
assumed temperature of 0.1\,K. If we now make use of~(\ref{vari}) then
the conclusion is that an integration time of about 3$\times$10$^{10}$
years is required to detect such a small temperature excess as the
optimized antenna permits\ldots

It can be argued that stochastic background signals will actually
be \emph{hotter} than the just described precision limit. However
nucleosynthesis constraints, for instance, set bounds on the spectral
density of GWs to values below 3$\times$10$^{-24}$\,Hz$^{-1/2}$ at
frequencies in the 1\,kHz range~\cite{Maggiore}. This translates into a
GW bath temperature of fractions of a micro-Kelvin, for whose accurate
estimation integration times of the order of several 10\,000 years
would be required. The use of \emph{fast variance estimation}
techniques~\cite{ppp} may improve on this by between one and two
orders of magnitude, but even so one would still be in the range
of 100 to 1000 years of integration time ---an absurd figure.

There is also little hope that GW backgrounds originating in
astrophysical sources, such as e.g.\ supernovae~\cite{valeria}, have
an effective temperature above 10$^{-7}$\,K or so~\cite{Maggiore}.
\end{itemize}

Altogether then the \emph{multimode} capabilities of a spherical
GW antenna do not seem to offer the practical possibility to make
reliable, \emph{single detector} measurements of the spectral density
of a background of stochastic GWs. An exception to this however happens
when some \emph{a priori} information on the actual form of the spectral
function is known ahead of time (\emph{template})~\cite{uwa}. If this is
the case then \emph{optimum filter} techniques can be used, of which a
wealth is available in the literature~\cite{oscar}.

Of course this is not specific to \emph{spherical} detectors ---in fact
any single, unimodal GW antenna (such as bars or interferometers) can
make advantageous use of the above techniques. Spheres however are more
efficient in two senses: first they are \emph{omnidirectional}, and
second they can provide up to \emph{five independent estimates} of
the spectral density of the GW background, as follows from
equation~(\ref{diacorr}): each \emph{quadrupole channel} $y^{(m)}(t)$
can be used to produce one such independent estimate. On average, one
gets a factor of 15/8 because of omnidirectionality, times a factor of~5
for quadrupole channels, which make up for~75/8, one order of magnitude
(in energy) better than bars and interferometers. Let us however stress
that this obviously applies only within the frequency band of sensitivity
of each detector; in the case of a dual sphere, even this is quite
competitive, see~\cite{dual}.

\section{Conclusion \label{sec.8}}

In this paper we have investigated in detail how a spherical detector
interacts with a background of gravitational waves, characterized by a
frequency dependent \emph{spectral energy density} function, $S_h(\w)$.
In particular, we have found the generic response functions for a device
with an arbitrary number of motion sensors, as well as their correlation
matrix, both of the system readouts and for the five quadrupole channels
---\emph{mode channels} for certain specific sensor layout geometries.

We have discovered that there is no \emph{distinctive pattern} in that
correlation matrix, relative to the local noise induced pattern, except
that the latter involves \emph{all} the antenna's oscillation eigenmodes
rather than just the monopole/quadrupole harmonics, characteristic of
generic GWs. This difference proves however insufficient to efficiently
filter the stochastic GWs from random local disturbances.

We are thus led to conclude that, in the absence of some \emph{a priori}
knowledge about the GW spectral density $S_h(\w)$, the multimode character
of a spherical detector does not offer a useful alternative to the
traditional cross-correlation between two (or more) detectors as a
recipe to get rid of local noise effects. However, if such \emph{a priori}
knowledge is available then the spherical detector naturally yields one
order of magnitude better performance (in energy) than single readout
devices, such as bar and interferometric GW detectors.

\section*{Acknowledgments}

We are indebted with Antonello Ortolan and David Blair for helpful and
very illuminating discussions. JAL acknowledges support received from
the Spanish Ministry of Science, contract number BFM2000-0604. AM thanks
Generalitat de Catalunya and Universitat de Barcelona for a contract.

\section*{Appendix A}

A suitable representation for the matrices $E^{(00)}_{ij}$ and
$E^{(2m)}_{ij}$ used in equation~(\ref{eq.monqua}) is the following:

\bea
& & E^{(00)}_{ij} = \left\{\frac{1}{4\pi}\right\}^{\frac{1}{2}} 
     \pmatrix{1&0&0 \cr 0&1&0  \cr 0&0&1} \hspace{2.4 cm}   
    E^{(20)}_{ij} = \left\{\frac{5}{16\pi}\right\}^{\frac{1}{2}}
     \pmatrix{-1&0&0 \cr 0&-1&0 \cr 0&0&2} \nonumber \\[1 em]
& & E^{(2\pm1)}_{ij} = \left\{ \frac{15}{32\pi}\right\}^{\frac{1}{2}}
     \pmatrix{0&0&\mp1 \cr 0&0&-i \cr\mp1&-i&0} \hspace{1.0 cm}
    E^{(2\pm2)}_{ij} = \left\{\frac{15}{32\pi}\right\}^{\frac{1}{2}}
     \pmatrix{1&\pm i&0 \cr \pm i&-1&0 \cr 0&0&0} \hspace{1.5 cm}
\eea
having the properties:

\beq
E_{ij}^{(00)}\,\Omega_i\Omega_j = Y_{00}(\dir) \hspace{1 cm} , \hspace{1 cm}
E_{ij}^{(2m)}\,\Omega_i\Omega_j = Y_{2m}(\dir)
\eeq
where $\dir$ is a unit vector, and $Y_{lm}(\dir)$ are spherical harmonics.

\end{document}